\documentclass[sn-mathphys-num]{sn-jnl}

\usepackage{graphicx}%
\usepackage{multirow}%
\usepackage{amsmath,amssymb,amsfonts}%
\usepackage{amsthm}%
\usepackage{mathrsfs}%
\usepackage[title]{appendix}%
\usepackage{xcolor}%
\usepackage{textcomp}%
\usepackage{manyfoot}%
\usepackage{booktabs}%

\usepackage{listings}%

\usepackage{amsmath,amsfonts,amsbsy,amssymb,mathtools,amsthm,tabulary,graphicx,times,caption,fancyhdr, mathrsfs, pgf, epsfig, epstopdf, lscape, subfig,booktabs, enumitem,lipsum,tkz-graph,scalerel,stackengine,setspace,kbordermatrix,algpseudocode, algorithm, accents, verbatim, dsfont}

\newlength{\depthofsumsign}
\GraphInit[vstyle = Shade]
\tikzset{
	LabelStyle/.style = { rectangle, rounded corners, draw,
		minimum width = 2em, fill = yellow!50,
		text = red, font = \Large\bfseries },
	VertexStyle/.append style = { inner sep=5pt,
		font = \Large\bfseries},
	EdgeStyle/.append style = {->, bend left} }

\tikzset{
	mynode/.style={
		draw,
		thick,
		anchor=south west,
		minimum width=2cm,
		minimum height=1.3cm,
		align=center, 
		inner sep=0.2cm,
		outer sep=0,
		rectangle split, 
		rectangle split parts=2,
		rectangle split draw splits=false},
	reverseclip/.style={
		insert path={(current page.north east) --
			(current page.south east) --
			(current page.south west) --
			(current page.north west) --
			(current page.north east)}
	}
}

\tikzstyle{io} = [trapezium, trapezium left angle=70, trapezium right angle=110, minimum width=3cm, minimum height=0.5cm, text centered, draw=black, fill=blue!30]
\tikzstyle{process} = [rectangle, minimum width=3cm, minimum height=1cm, text centered, draw=black, fill=orange!30]
\tikzstyle{process2} = [rectangle, minimum width=3cm, minimum height=1cm, text centered, draw=black, fill=green!30]
\tikzstyle{process3} = [rectangle, minimum width=2cm, minimum height=1cm, text centered, draw=black, fill=yellow!30]

\DeclareFontFamily{U}{mathx}{\hyphenchar\font45}
\DeclareFontShape{U}{mathx}{m}{n}{
	<-6> mathx5 <6-7> mathx6 <7-8> matha7
	<8-9> mathx8 <9-10> mathx9
	<10-12> mathx10 <12-> mathx12
}{}
\DeclareSymbolFont{mathx}{U}{mathx}{m}{n}
\DeclareMathSymbol{\bigominus}{\mathop}{mathx}{"C1}

\DeclareMathAlphabet\mathbfcal{OMS}{cmsy}{b}{n}

\stackMath
\newcommand\reallywidehat[1]{%
	\savestack{\tmpbox}{\stretchto{%
			\scaleto{%
				\scalerel*[\widthof{\ensuremath{#1}}]{\kern-.6pt\bigwedge\kern-.6pt}%
				{\rule[-\textheight/2]{1ex}{\textheight}}
			}{\textheight}%
		}{0.5ex}}%
	\stackon[1pt]{#1}{\tmpbox}%
}

\theoremstyle{thmstyleone}%
\theoremstyle{thmstyletwo}%

\theoremstyle{thmstylethree}%
\newtheorem{definition}{Definition}%

\begin{document}

\title{Multiscale Causal Analysis of Market Efficiency via News Uncertainty Networks and \\the Financial Chaos Index}


\author*[1]{\fnm{Masoud} \sur{Ataei}}\email{masoud.ataei@utoronto.ca}

\affil[1]{\normalsize\orgdiv{Department of Mathematical and Computational Sciences},\\ \orgname{University of Toronto}, \state{Ontario}, \country{Canada}}



\abstract{
This study evaluates the scale-dependent informational efficiency of stock markets using the Financial Chaos Index, a tensor-eigenvalue-based measure of realized volatility. Incorporating Granger causality and network-theoretic analysis across a range of economic, policy, and news-based uncertainty indices, we assess whether public information is efficiently incorporated into asset price fluctuations. Based on a 34-year time period from 1990 to 2023, at the daily frequency, the semi-strong form of the Efficient Market Hypothesis is rejected at the 1\% level of significance, indicating that asset price changes respond predictably to lagged news-based uncertainty. In contrast, at the monthly frequency, such predictive structure largely vanishes, supporting informational efficiency at coarser temporal resolutions. A structural analysis of the Granger causality network reveals that fiscal and monetary policy uncertainties act as core initiators of systemic volatility, while peripheral indices, such as those related to healthcare and consumer prices, serve as latent bridges that become activated under crisis conditions. These findings underscore the role of time-scale decomposition and structural asymmetries in diagnosing market inefficiencies and mapping the propagation of macro-financial uncertainty.
 }




\maketitle

\newpage
\section{Introduction}

Understanding how information diffuses into financial markets remains a cornerstone of empirical finance. Central to this inquiry is the Efficient Market Hypothesis (EMH), which posits that asset prices reflect all available information. The EMH has been extensively debated in both theoretical and empirical contexts~\citep{degutis2014efficient, gabriela2015efficient, lehmann1990fads, stein1989efficient, shi2018should}. Recent advances in data availability and statistical modeling have allowed for more detailed evaluations of this hypothesis across different temporal resolutions and information channels.

In this study, we assess market efficiency by incorporating the \textit{Financial Chaos Index} (FCIX), a high-dimensional, tensor-eigenvalue-based measure that captures mutual variations in multivariate asset prices~\citep{ataei2021theory}. Unlike conventional scalar metrics of volatility, the FCIX is derived from a multilinear algebraic framework that encodes higher-order dependencies and systemic co-movements, thereby offering a geometrically enriched view of realized volatility and its latent structure.

Our empirical framework evaluates the FCIX jointly with the CBOE Volatility Index (VIX), a measure of implied volatility, and a broad array of sector-specific news-based uncertainty indices, including the Economic Policy Uncertainty (EPU) index~\citep{baker2016measuring}, the Equity Market Volatility (EMV) tracker~\citep{baker2019policy}, and eleven additional indices capturing fiscal, regulatory, healthcare, and geopolitical dimensions of uncertainty. The influence of news shocks on asset prices is well-documented, particularly in the presence of macroeconomic or policy disruptions that elevate perceived risk and reduce forecast accuracy~\citep{al2019economic, megaritis2021stock, brogaard2022moves}.

To examine causal linkages among these variables, we adopt an extended Granger causality framework~\citep{schiatti2015extended}, which jointly captures both strictly lagged and instantaneous dependencies. Accounting for instantaneous relations is crucial, as omitting them can confound lagged causality estimates~\citep{faes2014assessing}. Our implementation is carried out within a vector autoregressive (VAR) system encompassing realized and implied volatility, as well as the network of economic- and policy-relevant uncertainty signals.

Our study interrogates the \emph{semi-strong} form of the EMH, which asserts that publicly available news is immediately reflected in asset prices. Within our framework, this formulation is assessed through the presence (or absence) of strictly lagged Granger causal effects from news-based uncertainty indices to the FCIX. The results, based on a 34-year time period from 1990 to 2023, indicate that at the daily frequency, the semi-strong form of EMH is statistically rejected at the \(1\%\) level, revealing that asset price volatility is not fully responsive to public information in real time. However, when re-evaluated at the monthly frequency, we find support for market efficiency, consistent with the notion that information frictions and microstructure noise are attenuated over longer temporal horizons. These findings echo a growing literature on time-varying efficiency, emphasizing the need to consider both the granularity of information flows and the structure of uncertainty propagation~\citep{guidi2011weak,woo2020review, charfeddine2016time,ito2016evolution}.

In addition to hypothesis testing, we apply network-theoretic tools to the Granger causal graph linking macroeconomic and policy-related indicators. Centrality measures, path lengths, and node-specific scores such as betweenness and bridging coefficients are used to infer the transmission pathways through which uncertainty diffuses. Fiscal and monetary policy indices emerge as dominant systemic drivers, while healthcare and consumer price uncertainties act as strategic bridges under particular market regimes, especially during the COVID-19 pandemic. These structural insights advance our understanding of how different sources of uncertainty jointly influence realized market dynamics.

The remainder of the paper is structured as follows. Section~\ref{Sec:FCIX} reviews the construction of the FCIX and presents its properties. Section~\ref{Section_Forces} analyzes interactions between FCIX, VIX, and daily news-based uncertainty indices using a Granger causality framework, with particular attention to implications for market efficiency. Section~\ref{Section_Eff_Market} broadens the analysis to monthly data, examining the structural propagation of macroeconomic and policy-related uncertainty through causal networks and graph-theoretic measures. Finally, Section~\ref{Sec:Conclusion} concludes with a summary of the main findings and a discussion of their implications for informational efficiency and systemic volatility in financial markets.

\subsection{Preliminaries}

Throughout the paper, scalar quantities, vectors, and matrices are denoted by lowercase lightface (e.g., $x$), lowercase boldface (e.g., $\mathbf{x}$), and uppercase boldface (e.g., $\mathbf{X}$) letters, respectively, with all vectors presumed to be column vectors unless otherwise stated. Boldface calligraphic letters (e.g., $\boldsymbol{\mathcal{X}}$) denote tensors, multidimensional arrays that generalize matrices to higher-order structures. We use $\mathbb{R}_{\geq 0}$ and $\mathbb{R}_{>0}$ to denote the sets of nonnegative and strictly positive real numbers, respectively.

A tensor of order $N$ is an element of the Cartesian space $\mathbb{R}^{I_1 \times I_2 \times \cdots \times I_N}$, where $I_n$ denotes the dimension along the $n$th mode, and the total number of entries is given by $K = \prod_{n=1}^N I_n$. Analogous to rows and columns in matrices, a \emph{mode-$n$ fiber} of a tensor is obtained by fixing all indices except the $n$th, while fixing all but two indices yields a \emph{slice}, i.e., a matrix cross-section of the tensor. 

The Frobenius norm of a tensor $\boldsymbol{\mathcal{X}}$ is defined as
\begin{equation}
	\|\boldsymbol{\mathcal{X}}\|_F = \sqrt{\langle \boldsymbol{\mathcal{X}}, \boldsymbol{\mathcal{X}} \rangle} = \sqrt{ \sum_{i_1=1}^{I_1} \sum_{i_2=1}^{I_2} \cdots \sum_{i_N=1}^{I_N} x_{i_1 i_2 \ldots i_N}^2 },
\end{equation}
generalizing the Euclidean norm for vectors and the matrix Frobenius norm to higher-order settings.

A foundational tool in tensor analysis is the \emph{polyadic decomposition}, which approximates a tensor by a sum of rank-one components. For a given positive integer $R$, the rank-$R$ approximation of a tensor $\boldsymbol{\mathcal{X}}$ is given by
\begin{equation}
	\label{Eq:CPD}
	\boldsymbol{\mathcal{X}} \approx \widehat{\boldsymbol{\mathcal{X}}} = \sum_{r=1}^{R} \mathbf{a}_r \circ \mathbf{b}_r \circ \mathbf{c}_r,
\end{equation}
where $\circ$ denotes the outer product between vectors. If $R$ is the minimal number of rank-one terms that exactly reconstruct the tensor, the decomposition is referred to as the \emph{canonical polyadic decomposition} (CPD), and $R$ is the CP rank of $\boldsymbol{\mathcal{X}}$.

A tensor $\boldsymbol{\mathcal{X}}$ is said to be of rank one if it admits a factorization of the form
\begin{equation}
	\boldsymbol{\mathcal{X}} = \mathbf{a} \circ \mathbf{b} \circ \mathbf{c},
\end{equation}
where $\mathbf{a} \in \mathbb{R}^{I_1}$, $\mathbf{b} \in \mathbb{R}^{I_2}$, and $\mathbf{c} \in \mathbb{R}^{I_3}$. Each term in a polyadic decomposition can be interpreted as a latent directional component in the tensor structure, capturing underlying multilinear patterns among modes.

The tensor-based framework adopted in this paper incorporates these constructs to quantify systemic volatility in equity markets.

\section{Financial Chaos Index}
\label{Sec:FCIX}

The FCIX is a tensor-eigenvalue-based construct designed to quantify realized volatility in equity markets by capturing mutual fluctuations among asset prices~\citep{ataei2021theory}. In contrast to scalar volatility metrics, the FCIX incorporates higher-order tensorial representations to encode systemic co-movements, offering a geometrically enriched and temporally resolved characterization of market stress.


Let $\mathbf{r}^{(t)} \in \mathbb{R}_{>0}^N$ denote the vector of rates of return for $N$ assets at time $t \in \mathcal{T}$, where
\begin{equation}
	r^{(t)}_i = \frac{C^{(t)}_i}{C^{(t-1)}_i},
\end{equation}
and $C^{(t)}_i$ represents the adjusted closing price of asset $i$ at time $t$. Our empirical analysis covers $N = 811$ unique stocks that have appeared in the S\&P 500 index between January 1990 and December 2023, with complete price records across the time horizon.\footnote{Price data were sourced from the Center for Research in Security Prices (CRSP) via Wharton Research Data Services (WRDS).} The time index set $\mathcal{T}$ includes $8265$ daily and $408$ monthly observations.

At each $t$, we construct a \emph{reciprocal pairwise comparison matrix} (RPCM) $\mathbf{A}^{(t)} \in \mathbb{R}_{>0}^{N \times N}$ as follows:
\begin{equation}
	\mathbf{A}^{(t)} = \mathbf{r} \circ \left(\mathbf{r}^{-1}\right)^{\top},
\end{equation}
where element-wise inversion is denoted by $\mathbf{r}^{-1}$. Each entry $A^{(t)}_{ij}$ expresses the relative strength of asset $i$ over asset $j$ at time $t$, yielding a multiplicative structure often interpreted as an \emph{advantage matrix}~\citep{ataei2020time}.

\begin{figure}
	\centering
	\includegraphics[scale=0.2]{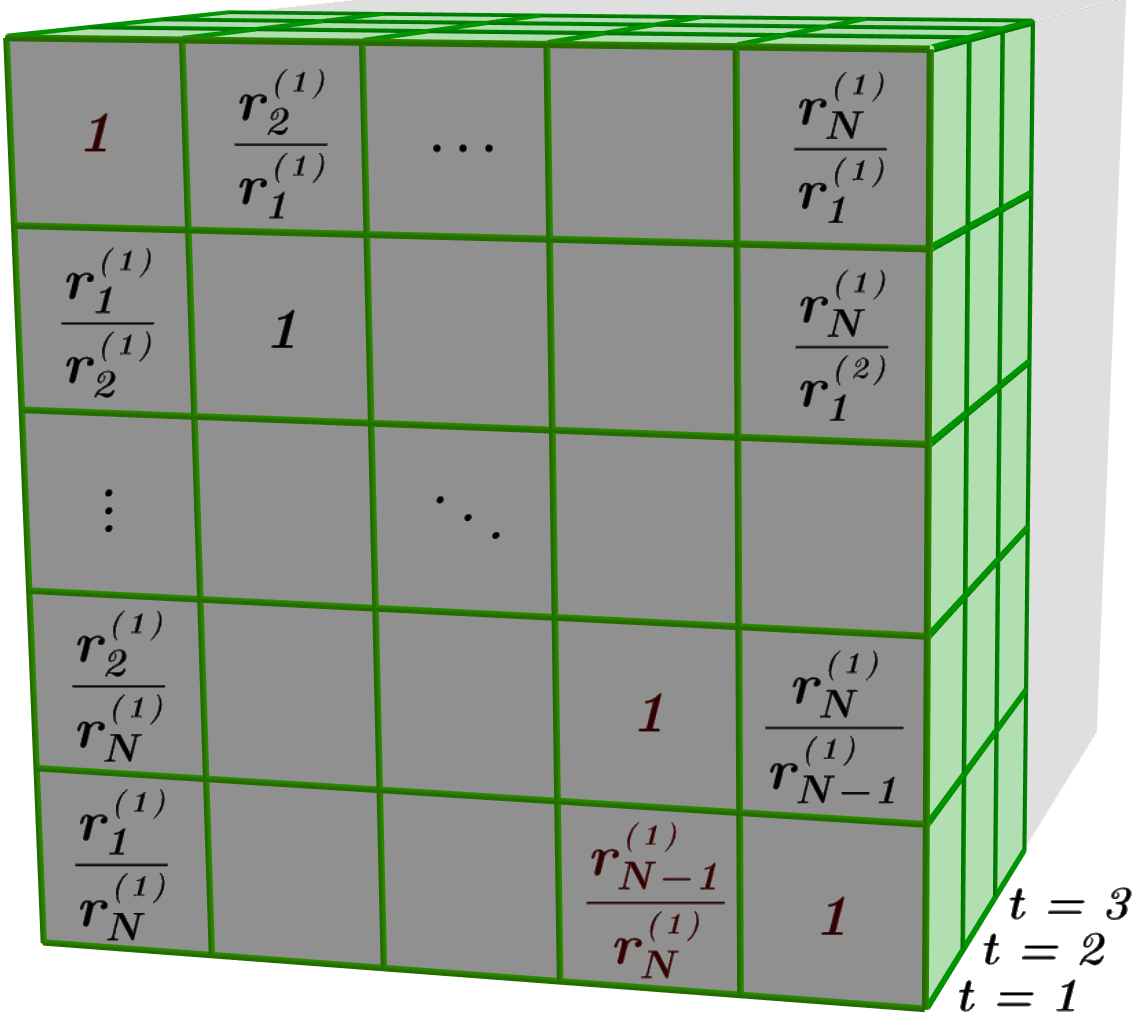}
	\caption{Schematic representation of the reciprocal pairwise comparison tensor (RPCT)~\citep{ataei2021theory}.}
	\label{Fig:RPCT}
\end{figure}

Stacking these matrices over time yields a third-order tensor $\boldsymbol{\mathcal{A}} \in \mathbb{R}_{>0}^{N \times N \times |\mathcal{T}|}$, termed the \textit{reciprocal pairwise comparison tensor} (RPCT), as depicted in Figure~\ref{Fig:RPCT}. The FCIX is then extracted by approximating $\boldsymbol{\mathcal{A}}$ via a rank-one tensor; i.e., we solve the following constrained optimization problem
\begin{equation}
	\begin{aligned}
		\min_{\tilde{\mathbf{x}}, \tilde{\mathbf{y}}, \tilde{\mathbf{z}}} \quad & \left\| \boldsymbol{\mathcal{A}} - \tilde{\mathbf{z}} \circ \left( \tilde{\mathbf{x}} \circ \tilde{\mathbf{y}}^\top \right) \right\|_F \\
		\text{s.t.} \quad & \tilde{z}_t \left( \tilde{\mathbf{x}} \circ \tilde{\mathbf{y}}^\top \right) > \mathbf{0}_N, \quad \forall t \in \mathcal{T}, \\
		& \tilde{\mathbf{x}}, \tilde{\mathbf{y}} \in \mathbb{R}_{>0}^N,\quad \tilde{\mathbf{z}} \in \mathbb{R}_{\geq 0}^{|\mathcal{T}|},
	\end{aligned}
\end{equation}
where $\mathbf{0}_N$ is the $N \times N$ zero matrix.

Once the approximation $\tilde{\boldsymbol{\mathcal{A}}} = \tilde{\mathbf{z}} \circ \left( \tilde{\mathbf{x}} \circ \tilde{\mathbf{y}}^\top \right)$ is obtained, the FCIX is defined at each time $t$ using the dominant eigenvalue of the frontal slice as follows:
\begin{equation}
	\operatorname{FCIX}(t) = \frac{\lambda_{\max}^{(t)} - N}{N-1},
\end{equation}
where $\lambda_{\max}^{(t)}$ denotes the largest eigenvalue of the matrix slice at time $t$. As shown in~\cite{ataei2021theory}, this construction aligns with the principal mode of systemic inconsistency and tracks latent volatility propagation across the asset universe. We further denote by $\mathrm{FCIX}_t$ the time series which takes on the observed FCIX values.

An important theoretical implication, established by~\citet{ataei2021theory}, is that the scaling vector $\tilde{\mathbf{z}}$ arising from the rank-one polyadic decomposition of the RPCT acts as a surrogate for the FCIX time series, modulo a positive rescaling, thereby linking the spectral geometry of the constituent RPCMs to the multilinear structure of the full tensor.

\begin{figure}
	\centering
	\includegraphics[width=\textwidth]{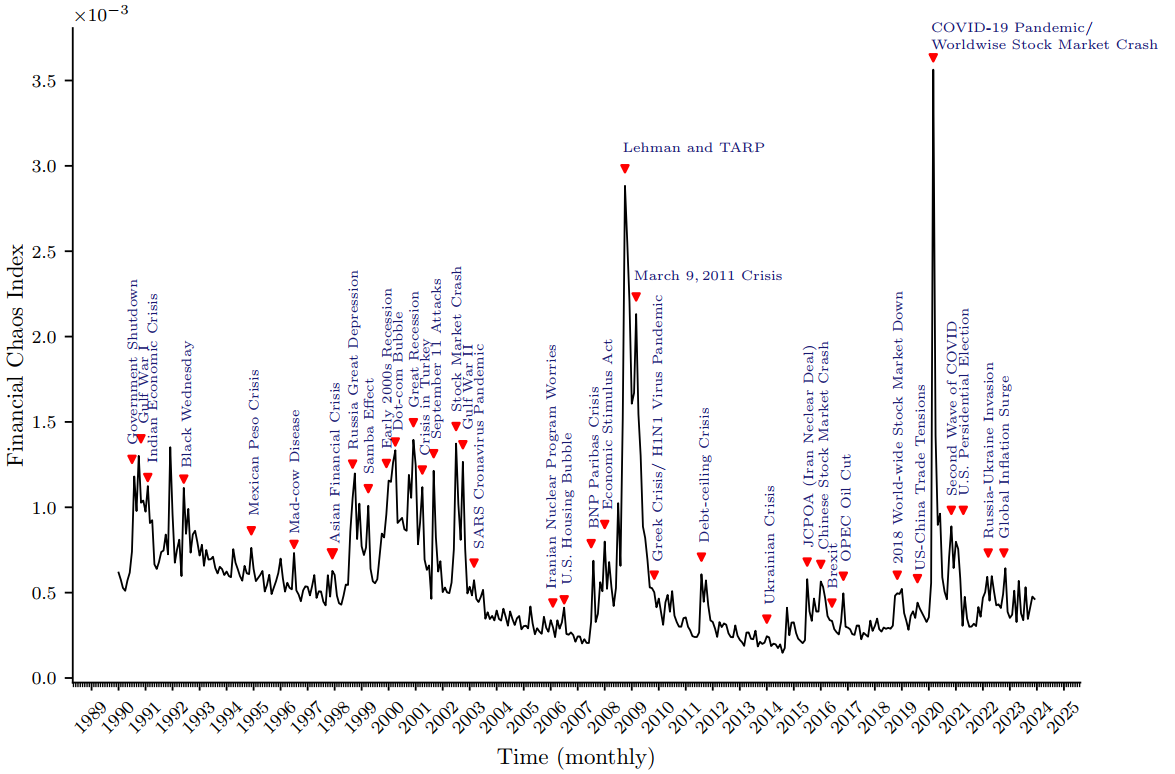} 
	\caption[Plot of the annotated monthly FCIX during January 1990--December 2023.]{Plot of the annotated monthly FCIX during January 1990--December 2023.}
	\label{Fig:MFCIX_Annotated} 
\end{figure}

A longitudinal analysis of monthly FCIX values from January 1990 to December 2023 reveals pronounced temporal heterogeneity. Figure~\ref{Fig:MFCIX_Annotated} highlights intermittent surges in FCIX aligned with historical episodes of financial stress, including the Dot-com collapse (2000--2002), the Global Financial Crisis (2008--2009), and the COVID-19 pandemic (2020). These events correspond to sharp excursions in FCIX, reflecting abrupt shifts in inter-asset correlations and systemic coherence.

Moreover, the empirical behavior of the FCIX is asymmetric: transitions into high-volatility regimes occur more abruptly than reversions, consistent with volatility clustering and asymmetric response to negative information. This dynamic asymmetry, where crisis-induced dislocations elevate the FCIX sharply, while recovery is gradual, reflects persistent structural fragility in financial networks.

Subsequently, the FCIX provides a tensor-based, geometrically interpretable measure of systemic volatility that captures higher-order dependencies among asset returns. Its spectral construction reveals sharp, asymmetric responses to financial crises. As a diagnostic tool, FCIX enables refined analysis of market fragility and regime shifts, supporting more detailed approaches to volatility modeling and systemic risk assessment.

%
%

\section{News Flow and Volatility Dynamics}
\label{Section_Forces}

In this section, we examine the interplay between realized and implied stock market volatilities, captured respectively by the FCIX and the VIX, and a range of macroeconomic and policy-related uncertainty measures. Our aim is to assess whether news-based indicators exhibit predictive power over realized volatility and, in doing so, to further test the semi-strong form of the EMH. To this end, we adopt the well-established framework of Granger causality, which operationalizes causality as improved out-of-sample forecastability within VAR systems.

This prediction-oriented approach is particularly appropriate in the context of market efficiency, where informational content is evaluated based on its capacity to improve volatility forecasts. In our setting, the VAR system includes FCIX, VIX, and a selection of uncertainty indices, permitting the joint analysis of informational flow across financial and news-based domains.

Two primary daily news-based uncertainty indices are considered: the EPU index~\citep{baker2016measuring} and the EMV tracker~\citep{baker2019policy}. The EPU index quantifies uncertainty by measuring the frequency of articles from ten major U.S. newspapers containing specific combinations of terms related to the economy, uncertainty, and policy institutions. In contrast, the EMV tracker is derived from a larger corpus of over a thousand news sources accessed via the Access World News' NewsBank service, capturing mentions of economic uncertainty in conjunction with terms directly referencing equity markets.

To account for the delayed market response to structural or macroeconomic shocks, we further incorporate monthly versions of the EPU and EMV indices, along with a comprehensive suite of category-specific monthly uncertainty indices. These include measures of uncertainty in monetary policy (MON), fiscal policy (FIS), tax policy (TAX), government spending (GOV), health care policy (HLTH), national security (SEC), entitlement programs (ENT), general regulation (GREG), financial regulation (FREG), trade policy (TRD), and consumer price inflation (CPI). These indices are constructed using refined keyword criteria similar to those employed for the EPU, allowing for thematic attribution of uncertainty across distinct policy domains~\citep{baker2016measuring}.

By integrating both daily and monthly sources of uncertainty into our Granger causality framework, we aim to systematically assess the informational transmission mechanisms underlying realized market volatility. This multi-resolution perspective enables us to identify the temporal and thematic structure of market reactions to news.

\subsection{Granger Causality Framework}

Let $(\Omega_t, \mathscr{F}_t, \mathbb{P})$ denote a probability space, where $\Omega_t$ represents the universe of market-relevant outcomes up to time $t$, and $\mathscr{F}_t$ is a $\sigma$-algebra representing the information actually known at time $t$, satisfying the filtration property $\mathscr{F}_1 \subseteq \mathscr{F}_2 \subseteq \cdots \subseteq \mathscr{F}_{|\mathcal{T}|}$. While $\Omega_t$ encompasses all possible latent factors, economic, financial and political, $\mathscr{F}_t$ restricts attention to measurable subsets, reflecting only the observable, agent-accessible information at time $t$.

Let $\{\mathbf{Y}_1, \mathbf{Y}_2, \dots, \mathbf{Y}_t\}$ denote a sequence of $K$-dimensional random vectors comprising a multivariate stochastic process $\{\mathbf{Y}_t: t \in \mathcal{T}\}$. Each vector $\mathbf{Y}_t: \Omega_t^K \rightarrow \mathcal{Y}_t \subseteq \mathbb{R}_{\geq 0}^K$ is a measurable map from $(\Omega_t^K, \bigotimes_{k=1}^K \mathscr{F}_{k,t})$ into the measurable space $(\mathcal{Y}_t, \mathscr{B}_{\mathcal{Y}_t})$, where $\mathscr{B}_{\mathcal{Y}_t}$ denotes the Borel $\sigma$-algebra on $\mathcal{Y}_t$. Here, the notation $\bigotimes_{k=1}^K \mathscr{F}_{k,t}$ refers to the product $\sigma$-algebra generated by the component processes, and $K$ represents the total number of time series under consideration.

Although the Cartesian product of $\sigma$-algebras is not itself a $\sigma$-algebra, we define the information set $\mathcal{J}_t := \bigotimes_{k=1}^K \mathscr{F}_{k,t}$ and interpret it as a Hilbert space spanned by the individual information components $\mathscr{F}_{k,t}$ under the $L^2$ inner product structure. For any two such Hilbert spaces $\mathcal{J}_t$ and $\mathcal{J}_t'$, we use $\mathcal{J}_t \bigoplus \mathcal{J}_t'$ to denote their orthogonal direct sum, and $\mathcal{J}_t \bigominus \mathscr{F}_{i,t}$ to indicate the removal of the subspace corresponding to $\mathbf{Y}_{i,t}$ from the information set.

Granger causality is founded on two principles: (i) a cause must precede its effect in time, and (ii) a cause should contribute predictive information about the effect not contained in other variables. Formally, $\mathbf{Y}_{i,t}$ is said to not Granger-cause $\mathbf{Y}_{j,t}$ with respect to $\mathcal{J}_t$ if the inclusion of $\mathscr{F}_{i,t}$ provides no improvement in the conditional forecast of $\mathbf{Y}_{j,t+h}$ beyond what is already inferred from $\mathcal{J}_t \bigominus \mathscr{F}_{i,t}$.

This notion is formalized below.\newline

\begin{definition}[Granger Causality]
	\label{Def:GC}
	Let $\mathbf{Y}_t$ be a $K$-dimensional stochastic process and let $\mathcal{J}_t$ denote the corresponding information set. Then, $\mathbf{Y}_{i,t}$ does not Granger-cause $\mathbf{Y}_{j,t}$ under the following scenarios
	\begin{enumerate}
		\item (Up to horizon $h > 0$): \\
		$\mathbf{Y}_{i,t} \overset{(h)}{\nrightarrow} \mathbf{Y}_{j,t} \quad \Leftrightarrow \quad \forall k \in \{1, \dots, h\}:\ 
		\mathbb{E}[ \mathbf{Y}_{j,t+k} | \mathcal{J}_t ] = \mathbb{E}[ \mathbf{Y}_{j,t+k} | \mathcal{J}_t \bigominus \mathscr{F}_{i,t} ]$.
		
		\item (At all horizons $h \geq 0$): \\
		$\mathbf{Y}_{i,t} \overset{(\infty)}{\nrightarrow} \mathbf{Y}_{j,t} \quad \Leftrightarrow \quad \forall h \geq 0:\ 
		\mathbb{E}[ \mathbf{Y}_{j,t+h} | \mathcal{J}_t ] = \mathbb{E}[ \mathbf{Y}_{j,t+h} | \mathcal{J}_t \bigominus \mathscr{F}_{i,t} ]$.
		
		\item (At fixed horizon $h \geq 0$): \\
		$\mathbf{Y}_{i,t} \overset{h}{\nrightarrow} \mathbf{Y}_{j,t} \quad \Leftrightarrow \quad 
		\mathbb{E}[ \mathbf{Y}_{j,t+h} | \mathcal{J}_t ] = \mathbb{E}[ \mathbf{Y}_{j,t+h} | \mathcal{J}_t \bigominus \mathscr{F}_{i,t} ]$.\newline
	\end{enumerate}
\end{definition}

When the horizon parameter $h$ is strictly positive, the causality relation reflects delayed or strictly causal effects. The special case $h = 0$ corresponds to \textit{instantaneous Granger causality}, where a dependence exists between the contemporaneous values $\mathbf{Y}_{i,t}$ and $\mathbf{Y}_{j,t}$ that is not attributable to their respective pasts. Such zero-lag effects can be interpreted as evidence of common latent shocks or simultaneous information assimilation.

Even though the literature on Granger causal analysis is extensive, most contributions focus solely on lagged (strictly causal) effects, overlooking the influence of contemporaneous interactions. However, ignoring such instantaneous relations can lead to biased or misattributed causal inferences, especially in high-frequency financial systems where news shocks and systemic adjustments may exert effects at zero lag. For a detailed discussion of this issue, see~\cite{faes2014assessing}. 

In the sequel, we adopt the framework of extended Granger causality (eGC) introduced by~\citet{schiatti2015extended}, which enables the joint analysis of strictly causal (\(h>0\)) and instantaneous (\(h=0\)) dependencies. This approach accommodates contemporaneous feedback structures and disentangles their effects from lagged dependencies.

To operationalize eGC, we consider a VAR model of order $p$ as follows:
\begin{equation}
	\label{Eq:VAR}
	\mathbf{Y}_t = \sum_{k=0}^p \mathbf{B}_k \mathbf{Y}_{t-k} + \boldsymbol{\mathcal{E}}_t,
\end{equation}
where $\mathbf{B}_k \in \mathbb{R}^{K \times K}$ denotes the coefficient matrix at lag $k$, and $\boldsymbol{\mathcal{E}}_t$ is the innovation process with $\mathbb{E}[\boldsymbol{\mathcal{E}}_t] = \mathbf{0}$ and $\mathrm{cov}(\boldsymbol{\mathcal{E}}_t) = \boldsymbol{\Sigma}$. The contemporaneous ($k=0$) component captures instantaneous influences among the variables.

To assess causality from $\mathbf{Y}_{i,t}$ to $\mathbf{Y}_{j,t}$, we estimate two models. The first is the \textit{unrestricted} model given by
\begin{equation}
	\label{Eq:Unrestricted}
	\mathbf{Y}_{j,t} = \sum_{k=0}^p \mathbf{B}_{j,k} \mathbf{Y}_{t-k} + \mathcal{E}_{j,t},
\end{equation}
where $\mathbf{B}_{j,k}$ is the $j$th row of $\mathbf{B}_k$, capturing the full influence of all $K$ variables at lag $k$. The second is the \textit{restricted} model given by
\begin{equation}
	\label{Eq:Rrestricted}
	\mathbf{Y}_{j,t} = \sum_{k=0}^p \mathbf{B}_{j,k}^\prime \mathbf{Y}_{t-k}^{(i)} + \mathcal{E}_{j,t}^\prime,
\end{equation}
where $\mathbf{Y}_{t-k}^{(i)}$ denotes the vector $\mathbf{Y}_{t-k}$ with the $i$th component removed, and $\mathbf{B}_{j,k}^\prime$ contains the adjusted coefficients. This restricted model thus encodes predictions made in the absence of $\mathbf{Y}_{i,t}$, i.e., based on $\mathcal{J}_t \bigominus \mathscr{F}_{i,t}$.

The strength of the causal influence from $\mathbf{Y}_{i,t}$ to $\mathbf{Y}_{j,t}$ is quantified using the log-likelihood ratio statistic given by
\begin{equation}
	\label{EQ:GC_Measure}
	\mathrm{eGC}_{i \to j} = \log \frac{|\Sigma_j|}{|\Sigma_j^\prime|},
\end{equation}
where $\Sigma_j = \mathrm{cov}(\mathcal{E}_{j,t})$ and $\Sigma_j^\prime = \mathrm{cov}(\mathcal{E}_{j,t}^\prime)$ are the covariance matrices of the unrestricted and restricted residuals, respectively. The eGC statistic is always non-negative and vanishes if and only if the exclusion of $\mathbf{Y}_{i,t}$ does not reduce predictive performance.

This formulation admits an interpretation grounded in information theory: $\mathrm{eGC}_{i \to j}$ measures the reduction in uncertainty about $\mathbf{Y}_{j,t}$ attributable to the inclusion of $\mathbf{Y}_{i,t}$ in the information set. When the residuals are assumed to be jointly Gaussian and the VAR is linear, it has been shown that Granger causality is equivalent to the transfer entropy defined in terms of conditional mutual information~\citep{barnett2012transfer}. Thus, eGC provides a bridge between time-series econometrics and the broader theory of dynamical information flow.

\begin{figure}
	\centering
	\begin{tikzpicture}[thick,scale=1, every node/.style={scale=0.6}, LabelStyle/.style = { rectangle, rounded corners, draw,minimum width = 2em, fill= yellow!50, text = black, font = \bfseries } , VertexStyle/.style ={draw, shape = circle,
			shading = ball, color=black, ball color = cyan!50!gray, minimum size = 50pt, font = \Large\bfseries} , EdgeStyle/.style =
		{->,bend left, thick, double = teal, double distance = 0.5pt} ]
		\SetGraphUnit{4}
		\Vertex{FCIX}
		\SO[Lpos=90,unit=3](FCIX){EPU}
		\NOEA[Lpos=90,unit=3](FCIX){VIX}
		\NOWE[Lpos=90,unit=3](FCIX){EMV}
		
		\Edge[label = $0.008$ ,style={bend left=0}](FCIX)(EMV)
		\Edge[label = $0.017$ ,style={bend left=0}](EMV)(VIX)
		\Edge[label = $0.035$ ,style={bend left=0}](VIX)(FCIX)
		\Edge[label = $0.128$ ,style={bend left=-30}](EMV)(EPU)
		
		\Edge[label = $0.003$, style={dashed,bend left=25}](FCIX)(VIX)
		\Edge[label = $0.007$, style={dashed,bend left=25}](EMV)(FCIX)
		\Edge[label = $0.003$, style={dashed,bend left=0}](FCIX)(EPU)
		\Edge[label = $0.015$, style={dashed,bend left=30}](VIX)(EPU)
		\Edge[label = $0.002$, style={dashed,bend left=0}](EPU)(VIX)
		\Edge[label = $0.004$, style={dashed,bend left=0}](EPU)(EMV)
		\Edge[label = $0.023$, style={dashed,bend left=-30}](VIX)(EMV)
		
		\Loop[dist=1.5cm,dir=NOEA,style={dashed,thick,->},label = $0.557$](FCIX.west)
		\Loop[dist=2cm,dir=SOWE,style={dashed,thick,->},label = $2.884$](VIX.east)
		\Loop[dist=2cm,dir=NOWE,style={dashed,thick,->},label = $0.228$](EMV.west)
		\Loop[dist=2cm,dir=SO,style={dashed,thick,->},label = $0.419$](EPU.east)
	\end{tikzpicture}
	\caption[Diagram of extended Granger causal relations among daily time series.]{Diagram of extended Granger causal relations among daily time series  at $1\%$ level of significance during January 1990-December 2023. Instantaneous and lagged relations are indicated using solid and dashed lines, respectively. The amount of extended GC measure \eqref{EQ:GC_Measure} is also presented for each relation.}
	\label{fig:GC_Daily}
\end{figure}
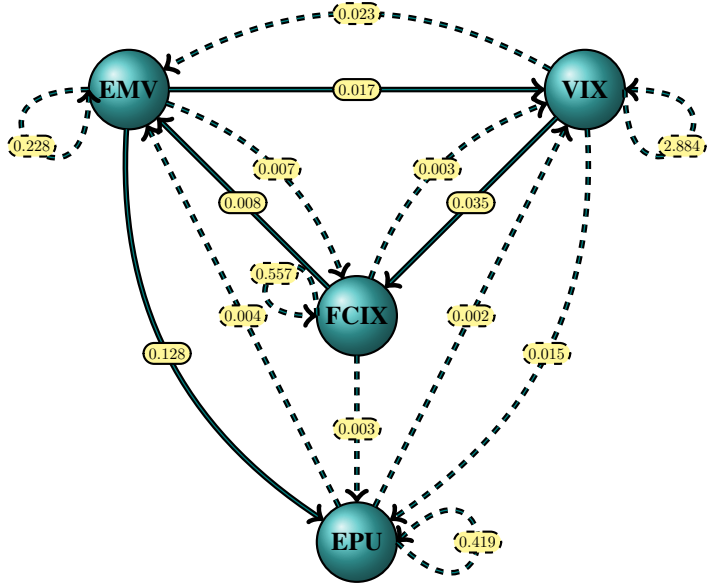

Figure~\ref{fig:GC_Daily} presents the eGC relations among the daily series $\left(\mathrm{FCIX}_t, \mathrm{VIX}_t, \mathrm{EMV}_t, \mathrm{EPU}_t\right)$ over the period January 1990 to December 2023. The diagram distinguishes between instantaneous (solid arrows) and lagged (dashed arrows) causal influences, with edge weights indicating the corresponding eGC magnitudes at the $1\%$ level of significance.

The most prominent feature of the causal structure is the presence of a unidirectional instantaneous cycle given by
\[
\mathrm{FCIX}_t \rightarrow \mathrm{EMV}_t \rightarrow \mathrm{VIX}_t \rightarrow \mathrm{FCIX}_t,
\]
which forms a closed triad of feedback operating without time lag. These links suggest a tightly interdependent information system among realized volatility, media-based equity market sentiment, and implied volatility, with each node contemporaneously contributing to the dynamics of the others. This observation substantiates the earlier evidence (cf. Section~\ref{Sec:FCIX}) concerning the high-frequency transmission of volatility shocks in modern equity markets.

Among the instantaneous effects, the strongest is observed from VIX to FCIX ($0.035$), indicating that shifts in implied volatility significantly and immediately influence the evolution of realized systemic volatility. The direct influence from EMV to VIX ($0.017$) further supports the role of media-driven sentiment as a short-horizon predictor of market fear. FCIX also exhibits an instantaneous impact on EMV ($0.008$), closing the feedback loop and underscoring the reflexivity of volatility measures and investor attention.

EMV displays a notable lagged effect on EPU ($0.128$), suggesting that media representations of equity market volatility have longer-term implications for perceptions of economic policy uncertainty. Additionally, EMV exerts a lagged influence on FCIX ($0.007$), while VIX and FCIX themselves exhibit lagged impacts on EPU ($0.015$ and $0.003$, respectively). These linkages imply a propagation channel through which financial volatility influences macro-level uncertainty not only contemporaneously but also with temporal delay.

The EPU series, in turn, shows weak but significant lagged feedback to EMV ($0.004$) and VIX ($0.002$), completing a broader, lower-magnitude feedback network. This suggests that economic policy uncertainty plays a limited predictive role in shaping short-run financial volatility but remains influenced by financial market conditions.

Self-dependencies, depicted via dashed loops, are present for all nodes and offer insight into the persistence of each process. The VIX index exhibits the strongest self-dependence ($2.884$), reaffirming its known autocorrelation structure and slow decay of shocks; see e.g., \citep{ataei2021theory} for further details. FCIX ($0.557$) and EMV ($0.228$) also show moderate persistence, consistent with clustering in realized and sentiment-based volatility measures. EPU ($0.419$) shows similar memory, though less pronounced than VIX.

Overall, the network in Figure~\ref{fig:GC_Daily} reveals a high-density, directionally asymmetric information structure. The core triad among FCIX, VIX, and EMV operates at near-zero latency, while EPU primarily serves as a lagged receptor of volatility dynamics. These findings point to an informational hierarchy in which market-based uncertainty measures not only reflect but also partially anticipate shifts in macroeconomic sentiment. The structure offers compelling empirical support for a time-scale dependent interpretation of market efficiency, wherein news shocks are rapidly assimilated at high frequency, while their longer-term consequences unfold across slower-moving economic channels.

\subsection{Daily Test of Market Efficiency}

We now turn to the semi-strong form of the EMH, which posits that publicly available information, including economic and policy-related news, is instantaneously and fully reflected in asset prices. In this setting, any predictive relation from news-based uncertainty indices to realized volatility should manifest only contemporaneously; that is, there should be no strictly time-lagged Granger causal effects. The presence of such lagged effects would signal that market participants fail to incorporate available information promptly, contradicting the semi-strong form of EMH; see \citep{kirchgassner2012introduction} for a classical treatment.

To formalize this idea, we adapt the concept of informational efficiency using the FCIX, which serves as a multivariate proxy for realized volatility.\newline

\begin{definition}[Semi-strong EMH]
	Let $\mathcal{J}_t^{\mathtt{PRC}}$ and $\mathcal{J}_t^{\mathtt{NEWS}}$ denote the information sets corresponding to the historical asset price structure (as reflected by the FCIX) and news-based uncertainty indices, respectively. The market is said to be semi-strongly efficient with respect to the joint information set $\mathcal{J}_t^{\mathtt{PRC}} \bigoplus \mathcal{J}_t^{\mathtt{NEWS}}$ if the process $\mathrm{FCIX}_t$ is not strictly Granger-caused by any component of $\mathcal{J}_t^{\mathtt{NEWS}}$.\newline
\end{definition}

To test this condition empirically, we examine whether the daily news-based indices $\mathrm{EMV}_t$ and $\mathrm{EPU}_t$ Granger-cause $\mathrm{FCIX}_t$ through strictly lagged effects. Specifically, we test the multiple null hypothesis
\[
\mathrm{H}_{0,\mathtt{D}}^{\mathtt{NEWS}} = \left( \mathrm{EMV}_t \overset{h>0}{\nrightarrow} \mathrm{FCIX}_t \right) \cap \left( \mathrm{EPU}_t \overset{h>0}{\nrightarrow} \mathrm{FCIX}_t \right),
\]
where $\overset{h>0}{\nrightarrow}$ denotes the absence of strictly lagged Granger causality. The individual $p$-values associated with these tests are computed as $p_1 = 0.000$ for EMV and $p_2 = 0.036$ for EPU, based on 500 bootstrap replications.

Given the rejection of at least one component hypothesis at conventional significance levels, we apply the Bonferroni correction to adjust for multiple testing~\citep{bland1995multiple}. The null $\mathrm{H}_{0,\mathtt{D}}^{\mathtt{NEWS}}$ is rejected at significance level $\alpha$ if
\[
\min\{p_1, p_2\} \leq \frac{\alpha}{2}.
\]
For $\alpha = 0.01$, the inequality is satisfied, implying that the joint hypothesis is rejected at the $1\%$ level. This leads to the conclusion that the daily FCIX is significantly Granger-caused by at least one news-based index through a strictly time-lagged mechanism.

These results reveal a structural lag in the incorporation of news-based uncertainty into realized market volatility. In particular, the influence of EMV on FCIX appears to be both statistically robust and economically meaningful. While EPU also exhibits weak lagged causality, its effect is less pronounced. Crucially, these findings contradict the semi-strong form of the EMH, which requires that public information be assimilated into asset prices without delay.

It is worth noting that while VIX was not included in the above hypothesis formulation, it remains an essential component of the multivariate system. As discussed in \cite{ataei2021theory}, FCIX and VIX are cointegrated, implying that omission of VIX from the modeling framework could bias the detection of Granger effects or induce spurious causality.

In conclusion, the empirical evidence points toward a violation of semi-strong informational efficiency in daily financial markets. The presence of strictly time-lagged causal relations from news-based indices to FCIX implies that macroeconomic and policy-related news is not instantaneously absorbed into asset price volatility. This inefficiency highlights the existence of latent frictions in the transmission of public information to realized market behavior, with implications for both trading strategies and regulatory oversight.


\section{Uncertainty Networks in Financial Systems}
\label{Section_Eff_Market}

This section extends the empirical analysis by incorporating a broader array of economic, financial, policy, and geopolitical indicators that collectively shape the dynamics of stock market volatility. By transitioning from daily to monthly frequency, we aim to capture more persistent structural shocks that operate over longer time horizons and whose effects diffuse gradually across macro-financial domains. This higher aggregation level allows for a more reliable attribution of market uncertainty to underlying policy and sector-specific conditions.

\subsection{Macroeconomic Causal Structure}

\begin{figure}
	\centering
	\includegraphics[scale=0.35]{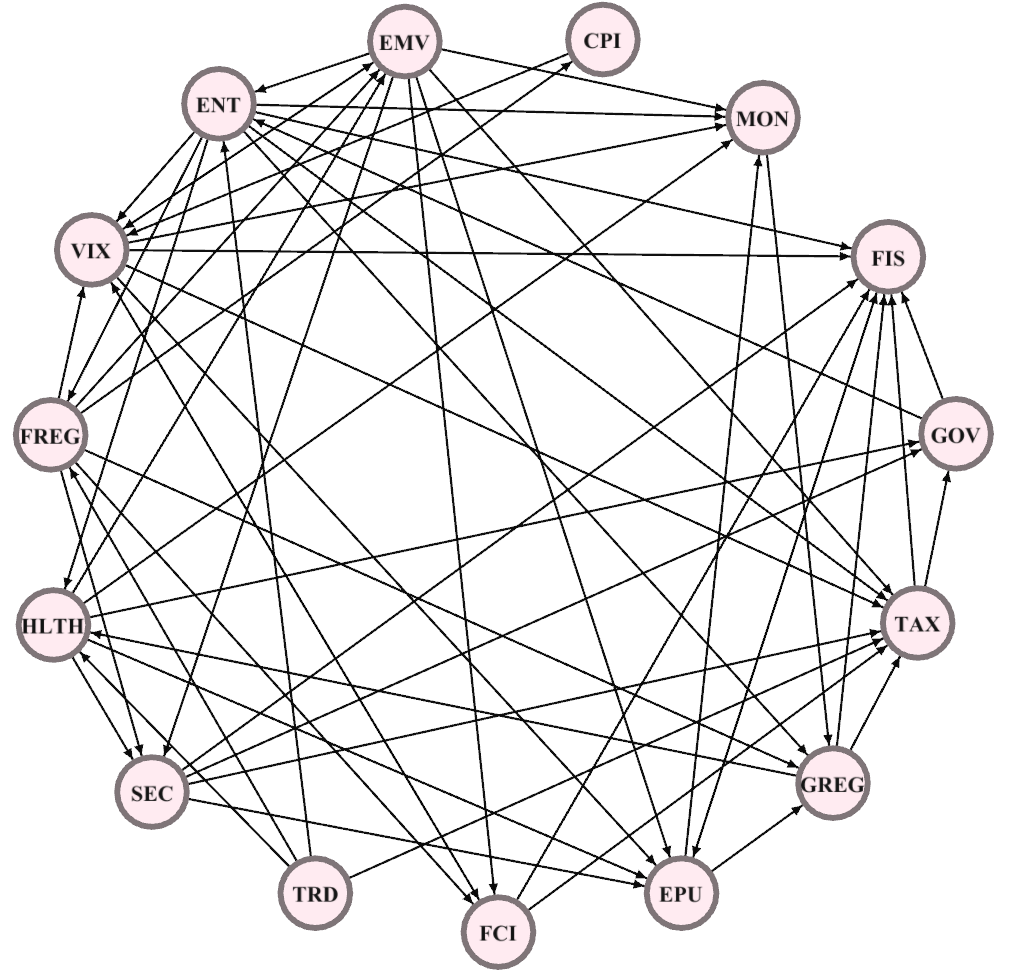}
	\caption[Diagram of extended Granger causal relations among monthly time series.]{Diagram of extended Granger causal relations among monthly time series at $1\%$ level of significance during January 1990-December 2023.}
	\label{fig:GC_Network}
\end{figure}

\begin{figure}
	\centering
	\includegraphics[scale=0.35]{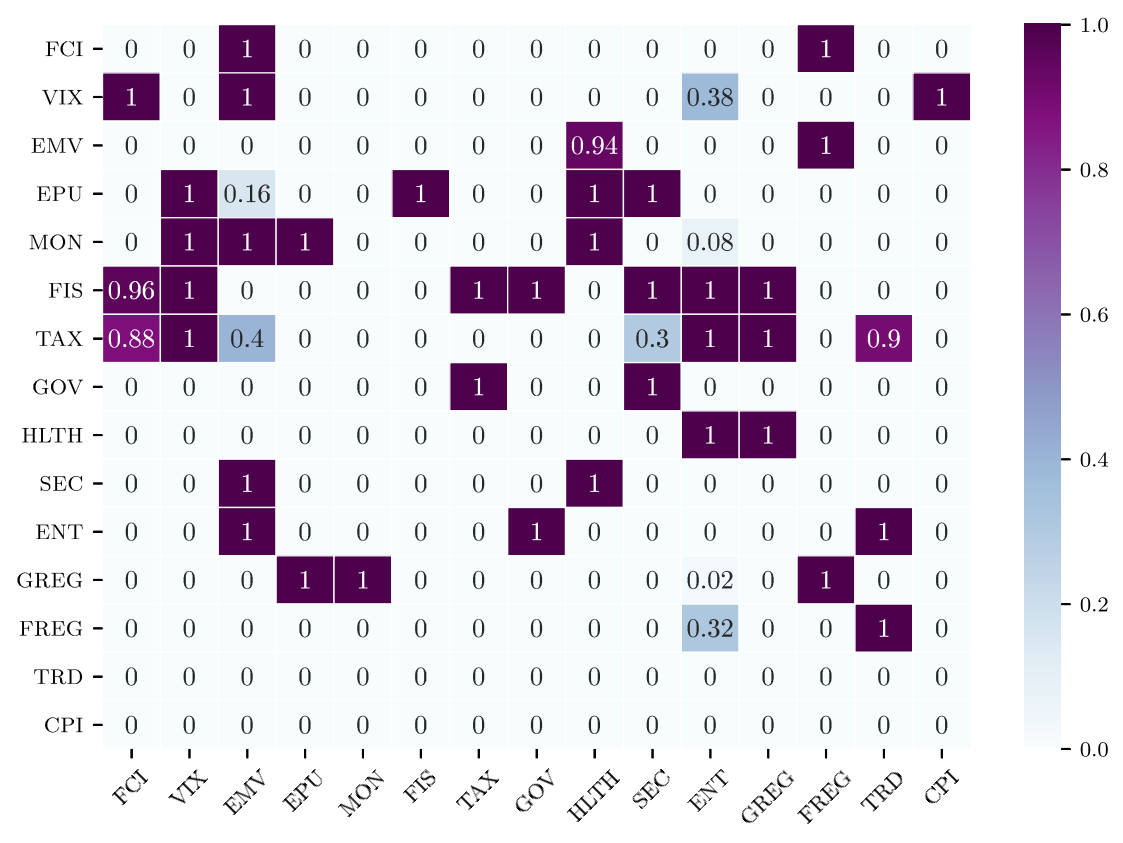}
	\caption[Matrix of empirical probability measures for instantaneous causal relations among monthly time series.]{Matrix of empirical probability measures for instantaneous causal relations among monthly time series at $1\%$ level of significance during January 1990-December 2023. The $(i,j)$-element represents the probability that the $j$-th time series Granger causes the $i$-th time series instantaneously.}
	\label{fig:heatmap_prob}
\end{figure}

\begin{figure}
	\centering
	\includegraphics[scale=0.35]{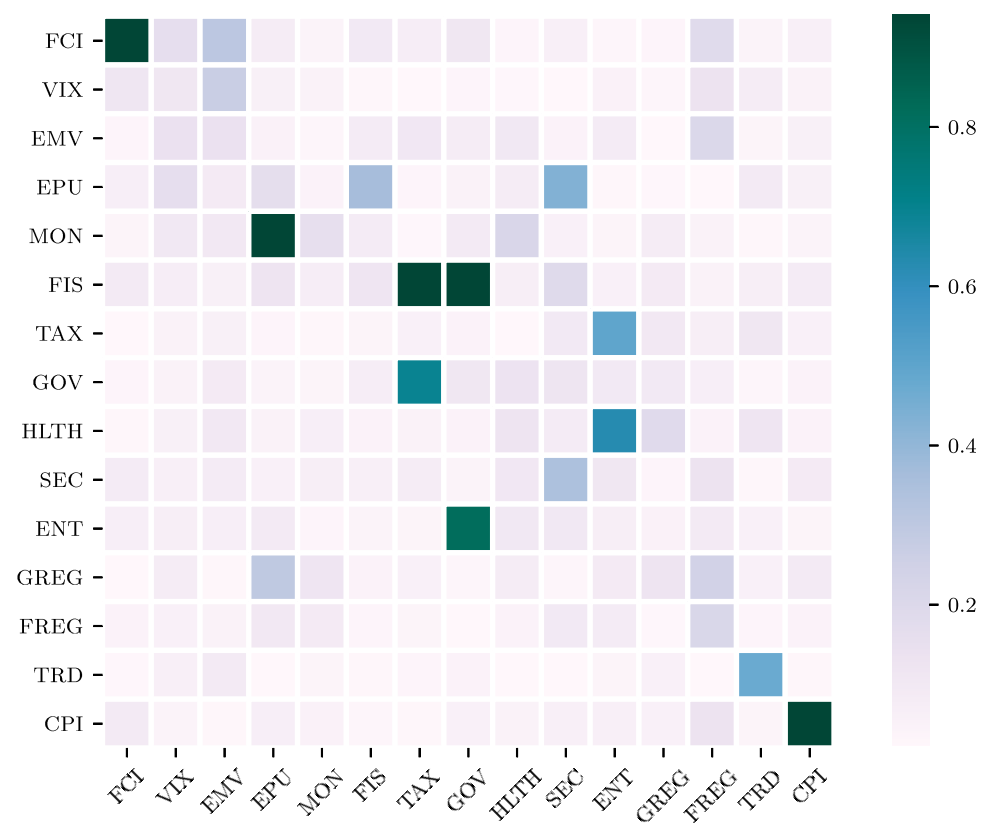}
	\caption[Heatmap of extended Granger causal measures among monthly time series.]{Heatmap of extended Granger causal measures among monthly time series during January 1990-December 2023. The $(i,j)$-element represents the measure from component $j$ to $i$, i.e., $\mathrm{eGC}_{j\to i}$.}
	\label{fig:heatmap_gc_measure}
\end{figure}

Let $\mathbf{Z}_t$ denote the multivariate time series capturing market-relevant uncertainty, defined as
\begin{equation*}
	\resizebox{1\hsize}{!}{$\mathbf{Z}_t = \left( \mathrm{FCIX}_t, \mathrm{VIX}_t, \mathrm{EMV}_t, \mathrm{EPU}_t, \mathrm{MON}_t, \mathrm{FIS}_t, \mathrm{TAX}_t, \mathrm{GOV}_t, \mathrm{HLTH}_t, \mathrm{SEC}_t, \mathrm{ENT}_t, \mathrm{GREG}_t, \mathrm{FREG}_t, \mathrm{TRD}_t, \mathrm{CPI}_t \right)^\top.$}
\end{equation*}
The extended Granger causality network and associated heatmaps in Figures~\ref{fig:GC_Network}--\ref{fig:heatmap_gc_measure} reveal a layered and asymmetrically connected structure of uncertainty propagation. At the core of the network lie monetary and fiscal policy uncertainties (MON and FIS), which emerge as dominant causal initiators. These variables exhibit extensive outbound links to both regulatory and market-based indices, underscoring their role as principal drivers of systemic volatility.

Several pathways of interest arise from the instantaneous causality matrix. Most notably, the relations $\mathrm{FIS} \to \mathrm{FCIX}$ (0.96) and $\mathrm{TAX} \to \mathrm{FCIX}$ (0.88) indicate that fiscal policy uncertainty exerts a nearly deterministic influence on realized market volatility. Similarly, the effect of financial regulation on implied volatility is captured by the moderately high measure $\mathrm{FREG} \to \mathrm{VIX}$ (0.32), suggesting that regulatory adjustments are rapidly internalized into investor sentiment.

Based on the empirical structure, we identify three functional layers within the system: \emph{initiators} (FIS, MON, TAX), which trigger directional shocks; \emph{transmitters} (EMV, GREG, FREG), which mediate and diffuse uncertainty across domains; and \emph{receivers} (FCIX, HLTH, SEC, CPI), which absorb external shocks with minimal outbound influence. The extended Granger causality matrix further validates this classification by revealing high column-wise densities for initiators and concentrated row-wise sensitivity among receivers.

Importantly, several multi-step causal chains emerge, highlighting the hierarchical and sectoral propagation of uncertainty. For instance, the pathway $\mathrm{MON} \to \mathrm{GREG} \to \mathrm{HLTH}$ captures how monetary policy shocks can manifest in downstream healthcare uncertainty via regulatory channels. Similarly, the sequence $\mathrm{FIS} \to \mathrm{TAX} \to \mathrm{CPI}$ reflects the transmission of fiscal instability into consumer price uncertainty.

Healthcare policy uncertainty ($\mathrm{HLTH}$), while not dominant in static centrality metrics, plays a latent bridging role that becomes more pronounced in crisis periods such as COVID-19. It receives input from regulatory and entitlement indices, but also propagates uncertainty to monetary and national security domains, thus functioning as a crisis-activated relay within the network.

Compared to the daily network of Section~\ref{Section_Forces}, where sentiment indices like EMV and VIX dominate, the monthly framework reveals a shift toward structural macro-policy variables. This temporal segmentation confirms that market sentiment is more volatile and reactive in the short term, whereas systemic uncertainty accumulates through slower-moving economic channels.

\subsection{Network Roles and Topology}
\label{Section:Graph_Analysis}

The extended Granger causality network shown in Figure~\ref{fig:GC_Network} may be further interrogated using graph-theoretic metrics to uncover latent structural properties governing uncertainty propagation. At a global level, the network exhibits a diameter of $7$, indicating that no more than seven steps are needed for information to flow across the most distant macroeconomic and policy-related indices in the system.

The average path length, computed as $2.326$, reflects the mean geodesic distance between all node pairs. This relatively small value, alongside the nontrivial diameter, suggests an efficient but non-uniform flow of information. The graph density, calculated as $0.314$, indicates that approximately 31\% of all possible directed connections are realized. This sparse but connected structure is typical in systems that balance complexity with containment.

Table~\ref{GC_Statistics} reports node-level diagnostics. Using the HITS algorithm of \cite{kleinberg1999authoritative}, we compute authority and hub scores. Nodes such as {FIS} and {TAX} dominate in authority, suggesting they are endpoints of significant uncertainty transmission. In contrast, {VIX}, {EMV}, and {ENT} have high hub scores, revealing their roles as primary broadcasters of information.

PageRank proximity scores affirm the systemic importance of {FIS} and {TAX}, confirming their gravitational pull in the network. Betweenness centrality isolates {ENT} as a pivotal bridge, facilitating connections between fiscal, regulatory, and volatility domains. Bridging coefficients, especially high for {CPI} and {GOV}, identify nodes that facilitate transitions between loosely connected regions, acting as structural bottlenecks or passageways.

Beyond these classical diagnostics, deeper inspection reveals additional role asymmetries. Nodes such as {TAX} and {FIS}, which attain high authority and PageRank but only moderate hub scores, serve primarily as accumulators of macro-financial shocks. Conversely, {ENT}, with the highest hub and betweenness scores but modest authority, emerges as a structural broadcaster, routing uncertainty between disparate policy sectors. These asymmetries reflect a functional split between systemic absorbers and propagators.

Nodes like {TRD} and {CPI}, although peripheral in terms of centrality, exhibit high bridging coefficients. Their topological placement allows them to function as conduits linking otherwise weakly coupled substructures. In particular, {CPI}, despite its minimal role in direct causal transmission?holds the highest bridging coefficient, suggesting a latent capacity to mediate trans, sectoral spillovers, particularly during inflationary episodes.

Regulatory indices such as {GREG} and {FREG} occupy middle ground: their modest authority and hub scores, coupled with high PageRank and intermediate betweenness, confirm their status as mid-network routers. They neither initiate nor absorb shocks but rather channel the effects of macro-policy variables toward sectoral domains like healthcare and consumer sentiment.

Healthcare uncertainty ({HLTH}) exhibits an especially distinct role. While not dominant in traditional centrality measures, its elevated betweenness centrality (0.191) suggests latent importance. This is consistent with its behavior during systemic shocks such as the COVID-19 pandemic, where it transitioned from a passive recipient to an active conduit of cross-domain volatility, bridging trade, regulation, and national security.

Taken together, the graph analysis reveals a finely layered architecture. Core drivers ({FIS}, {TAX}, {MON}) direct the flow of structural uncertainty; relays and broadcasters ({ENT}, {EMV}, {VIX}) shape its distribution; and bridges ({CPI}, {GOV}, {HLTH}) ensure connectivity across modular sectors. These patterns affirm that structural influence in financial networks is shaped not only by degree and centrality, but also by topological positioning, regime responsiveness, and transmission latency.

\begin{table}
	\centering
	\caption{Network statistics for monthly extended Granger causal relations.}
	\label{GC_Statistics}
	\begin{tabular}{lcccccc} 
		\toprule
		Node & Authority & Hub & \begin{tabular}[c]{@{}c@{}}Page Rank \\ Proximity\end{tabular} & \begin{tabular}[c]{@{}c@{}}Betweenness \\ Centrality\end{tabular} & \begin{tabular}[c]{@{}c@{}}Bridging \\ Coefficient\end{tabular} \\
		\midrule
		FCIX  & 0.229 & 0.255 & 0.368 & 0.003 & 0.208 \\
		VIX   & 0.300 & 0.422 & 0.424 & 0.128 & 0.084 \\
		EMV   & 0.193 & 0.426 & 0.412 & 0.156 & 0.079 \\
		EPU   & 0.304 & 0.211 & 0.583 & 0.109 & 0.123 \\
		MON   & 0.368 & 0.041 & 0.583 & 0.017 & 0.176 \\
		FIS   & 0.439 & 0.064 & 0.700 & 0.029 & 0.087 \\
		TAX   & 0.474 & 0.121 & 0.636 & 0.071 & 0.103 \\
		GOV   & 0.138 & 0.125 & 0.466 & 0.147 & 0.389 \\
		HLTH  & 0.183 & 0.251 & 0.424 & 0.191 & 0.091 \\
		SEC   & 0.194 & 0.285 & 0.412 & 0.018 & 0.124 \\
		ENT   & 0.158 & 0.440 & 0.400 & 0.246 & 0.085 \\
		GREG  & 0.197 & 0.230 & 0.519 & 0.198 & 0.133 \\
		FREG  & 0.134 & 0.245 & 0.304 & 0.114 & 0.088 \\
		TRD   & 0.000 & 0.199 & 0.000 & 0.000 & 0.288 \\
		CPI   & 0.051 & 0.063 & 0.250 & 0.000 & 0.567 \\
		\bottomrule
	\end{tabular}
\end{table}


\subsubsection{Monthly Test of Market Efficiency}
\label{Section_Eff_Market_1}

To investigate whether news-based uncertainty indexes strictly Granger cause $\mathrm{FCIX}_t$ over time-lagged horizons, we formulate the following family of null hypotheses
\begin{equation*}
	\mathrm{H}_{0,k}^{\mathtt{NEWS}}: \mathbf{Z}_{k,t} \overset{h>0}{\nrightarrow} \mathrm{FCIX}_t, \quad \forall k \in \{3,4,\dots,K\}.
\end{equation*}

Let $\mathcal{J}_{t,\mathtt{M}}^{\mathtt{PRC}} = \mathscr{F}_{1,t}$ and $\mathcal{J}_{t,\mathtt{M}}^{\mathtt{NEWS}} = \bigotimes_{k=3}^{K} \mathscr{F}_{k,t}$ denote the information sets corresponding to asset price changes and news-based uncertainty indexes, respectively, with $K=15$. The joint null hypothesis that no news-based index strictly Granger causes $\mathrm{FCIX}_t$ is given by
\begin{equation*}
	\mathrm{H}_{0,\mathtt{M}}^{\mathtt{NEWS}} = \bigcap_{k=3}^{K} \mathrm{H}_{0,k}^{\mathtt{NEWS}}.
\end{equation*}

Table~\ref{pValuesGC} reports the $p$-values associated with each time-lagged Granger test. Since the $p$-values are distributed across many components without strong concentration in any single variable, we adopt Fisher's combination test \citep{fisher1992statistical} to assess the joint hypothesis $\mathrm{H}_{0,\mathtt{M}}^{\mathtt{NEWS}}$. Under the null and assuming independence, the statistic
\begin{equation}
	-\log p_k \sim \mathrm{Exp}(1), \quad \forall k,
\end{equation}
implies that
\begin{equation}
	T_F = -2 \sum_{k=3}^{K} \log(p_k)
\end{equation}
follows a $\chi^2$ distribution with $2(K' - 2)$ degrees of freedom, where $K' \leq K$ accounts for any indices excluded due to instantaneous-only effects.

For the values in Table~\ref{pValuesGC}, we compute $T_F = 35.214$. When compared against the $\chi^2_{22}$ distribution, this yields a $p$-value of approximately $0.037$. As this exceeds the $1\%$ significance level, we fail to reject the global null $\mathrm{H}_{0,\mathtt{M}}^{\mathtt{NEWS}}$.

This outcome suggests that there is insufficient evidence to conclude that monthly news-based uncertainty indexes strictly Granger cause mutual asset price changes represented by $\mathrm{FCIX}$. Importantly, the strict causality formulation excludes contemporaneous effects, which were found to be significant in prior sections. Hence, this result does not contradict the presence of instantaneous transmission but rather confirms the absence of systematic lag structures.

From the perspective of the EMH, this finding is consistent with the semi-strong form of market efficiency. It supports the view that publicly available macroeconomic and policy-related news is rapidly and fully incorporated into asset prices at the monthly frequency, leaving no statistically reliable pathway for delayed predictive influence.

\begin{table}
	\centering
	\footnotesize
	\caption{Computed $p$-values for strictly causal relations among monthly time series.}
	\label{pValuesGC}
	\begin{tabular}{cc|cc|cc} 
		\toprule
		Relation & $p$-value & Relation & $p$-value & Relation & $p$-value \\
		\midrule
		$\mathrm{EPU}_t\overset{h>0}{\nrightarrow} \mathrm{FCIX}_t$ & 0.081 &
		$\mathrm{GOV}_t\overset{h>0}{\nrightarrow} \mathrm{FCIX}_t$ & 0.012 &
		$\mathrm{GREG}_t\overset{h>0}{\nrightarrow} \mathrm{FCIX}_t$ & 0.706 \\
		$\mathrm{MON}_t\overset{h>0}{\nrightarrow} \mathrm{FCIX}_t$ & 0.564 &
		$\mathrm{HLTH}_t\overset{h>0}{\nrightarrow} \mathrm{FCIX}_t$ & 0.681 &
		$\mathrm{TRD}_t\overset{h>0}{\nrightarrow} \mathrm{FCIX}_t$ & 0.528 \\
		$\mathrm{FIS}_t\overset{h>0}{\nrightarrow} \mathrm{FCIX}_t$ & 0.034 &
		$\mathrm{SEC}_t\overset{h>0}{\nrightarrow} \mathrm{FCIX}_t$ & 0.239 &
		$\mathrm{CPI}_t\overset{h>0}{\nrightarrow} \mathrm{FCIX}_t$ & 0.229 \\
		$\mathrm{TAX}_t\overset{h>0}{\nrightarrow} \mathrm{FCIX}_t$ & 0.114 &
		$\mathrm{ENT}_t\overset{h>0}{\nrightarrow} \mathrm{FCIX}_t$ & 0.788 &
		& \\
		\bottomrule
	\end{tabular}
\end{table}

\section{Conclusion}
\label{Sec:Conclusion}
This paper introduced a high-dimensional, information-theoretic framework for analyzing the dynamics of financial market uncertainty through the construction and empirical evaluation of the FCIX. By incorporating tensor-based representations of multivariate asset price variations, the FCIX provides a systematic measure of mutual information embedded in the joint structure of asset returns. Empirical assessments at both daily and monthly frequencies reveal several nontrivial statistical properties with implications for predictability and the efficient incorporation of information.

At the daily frequency, the detection of strictly time-lagged Granger causal relations from news-based uncertainty indices to the FCIX provides evidence against the semi-strong form of market efficiency at the 
1\% significance level, suggesting that public information is not instantaneously internalized into realized volatility. In contrast, the monthly analysis yields a markedly different picture. The statistical results support the semi-strong form of efficiency at the 1\% level. Extended Granger causality networks at this scale identify macroeconomic policy uncertainties, especially fiscal and monetary policy, as key drivers of systemic volatility. Complementary network-theoretic diagnostics highlight the structural importance of peripheral indices such as consumer price and healthcare uncertainties, which act as functional bridges during regime-switching episodes, including the COVID-19 pandemic.

Taken together, the findings underscore the necessity of frequency-aware modeling in capturing the layered and temporally heterogeneous structure of uncertainty transmission. The FCIX, in conjunction with extended Granger causality and graph-based diagnostics, offers a versatile and interpretable toolkit for assessing informational frictions and quantifying systemic vulnerability in high-dimensional financial environments.

Beyond its methodological contributions, the FCIX framework offers practical value for volatility forecasting, asset allocation, and macro-financial risk management. The identification of lagged and instantaneous causal structures can inform dynamic investment strategies, while the detection of central and bridging nodes within the uncertainty network supports the development of stress-testing protocols and adaptive hedging schemes. Future work may incorporate machine learning approaches for structural inference, or real-time filtering to dynamically track evolving uncertainty topologies in financial networks.

\newpage
\bibliography{references}


\begin{thebibliography}{23}
\ifx \bisbn   \undefined \def \bisbn  #1{ISBN #1}\fi
\ifx \binits  \undefined \def \binits#1{#1}\fi
\ifx \bauthor  \undefined \def \bauthor#1{#1}\fi
\ifx \batitle  \undefined \def \batitle#1{#1}\fi
\ifx \bjtitle  \undefined \def \bjtitle#1{#1}\fi
\ifx \bvolume  \undefined \def \bvolume#1{\textbf{#1}}\fi
\ifx \byear  \undefined \def \byear#1{#1}\fi
\ifx \bissue  \undefined \def \bissue#1{#1}\fi
\ifx \bfpage  \undefined \def \bfpage#1{#1}\fi
\ifx \blpage  \undefined \def \blpage #1{#1}\fi
\ifx \burl  \undefined \def \burl#1{\textsf{#1}}\fi
\ifx \doiurl  \undefined \def \doiurl#1{\url{https://doi.org/#1}}\fi
\ifx \betal  \undefined \def \betal{\textit{et al.}}\fi
\ifx \binstitute  \undefined \def \binstitute#1{#1}\fi
\ifx \binstitutionaled  \undefined \def \binstitutionaled#1{#1}\fi
\ifx \bctitle  \undefined \def \bctitle#1{#1}\fi
\ifx \beditor  \undefined \def \beditor#1{#1}\fi
\ifx \bpublisher  \undefined \def \bpublisher#1{#1}\fi
\ifx \bbtitle  \undefined \def \bbtitle#1{#1}\fi
\ifx \bedition  \undefined \def \bedition#1{#1}\fi
\ifx \bseriesno  \undefined \def \bseriesno#1{#1}\fi
\ifx \blocation  \undefined \def \blocation#1{#1}\fi
\ifx \bsertitle  \undefined \def \bsertitle#1{#1}\fi
\ifx \bsnm \undefined \def \bsnm#1{#1}\fi
\ifx \bsuffix \undefined \def \bsuffix#1{#1}\fi
\ifx \bparticle \undefined \def \bparticle#1{#1}\fi
\ifx \barticle \undefined \def \barticle#1{#1}\fi
\bibcommenthead
\ifx \bconfdate \undefined \def \bconfdate #1{#1}\fi
\ifx \botherref \undefined \def \botherref #1{#1}\fi
\ifx \url \undefined \def \url#1{\textsf{#1}}\fi
\ifx \bchapter \undefined \def \bchapter#1{#1}\fi
\ifx \bbook \undefined \def \bbook#1{#1}\fi
\ifx \bcomment \undefined \def \bcomment#1{#1}\fi
\ifx \oauthor \undefined \def \oauthor#1{#1}\fi
\ifx \citeauthoryear \undefined \def \citeauthoryear#1{#1}\fi
\ifx \endbibitem  \undefined \def \endbibitem {}\fi
\ifx \bconflocation  \undefined \def \bconflocation#1{#1}\fi
\ifx \arxivurl  \undefined \def \arxivurl#1{\textsf{#1}}\fi
\csname PreBibitemsHook\endcsname

\bibitem[\protect\citeauthoryear{Degutis and
  Novickyt{\.e}}{2014}]{degutis2014efficient}
\begin{barticle}
\bauthor{\bsnm{Degutis}, \binits{A.}},
\bauthor{\bsnm{Novickyt{\.e}}, \binits{L.}}:
\batitle{The efficient market hypothesis: A critical review of literature and
  methodology.}
\bjtitle{Ekonomika,}
\bvolume{93},
\bfpage{7}--\blpage{23}
(\byear{2014})
\end{barticle}
\endbibitem

\bibitem[\protect\citeauthoryear{Gabriela~{\u{g}}i{\.G}an}{2015}]{gabriela2015efficient}
\begin{barticle}
\bauthor{\bsnm{Gabriela~{\u{g}}i{\.G}an}, \binits{A.}}:
\batitle{The efficient market hypothesis: Review of specialized literature and
  empirical research.}
\bjtitle{Procedia Economics and Finance,}
\bvolume{32},
\bfpage{442}--\blpage{449}
(\byear{2015})
\end{barticle}
\endbibitem

\bibitem[\protect\citeauthoryear{Lehmann}{1990}]{lehmann1990fads}
\begin{barticle}
\bauthor{\bsnm{Lehmann}, \binits{B.N.}}:
\batitle{Fads, martingales, and market efficiency.}
\bjtitle{The Quarterly Journal of Economics,}
\bvolume{105}(\bissue{1}),
\bfpage{1}--\blpage{28}
(\byear{1990})
\end{barticle}
\endbibitem

\bibitem[\protect\citeauthoryear{Stein}{1989}]{stein1989efficient}
\begin{barticle}
\bauthor{\bsnm{Stein}, \binits{J.C.}}:
\batitle{Efficient capital markets, inefficient firms: A model of myopic
  corporate behavior.}
\bjtitle{The Quarterly Journal of Economics,}
\bvolume{104}(\bissue{4}),
\bfpage{655}--\blpage{669}
(\byear{1989})
\end{barticle}
\endbibitem

\bibitem[\protect\citeauthoryear{Shi and Delacroix}{2018}]{shi2018should}
\begin{barticle}
\bauthor{\bsnm{Shi}, \binits{S.}},
\bauthor{\bsnm{Delacroix}, \binits{A.}}:
\batitle{Should buyers or sellers organize trade in a frictional market?.}
\bjtitle{The Quarterly Journal of Economics,}
\bvolume{133}(\bissue{4}),
\bfpage{2171}--\blpage{2214}
(\byear{2018})
\end{barticle}
\endbibitem

\bibitem[\protect\citeauthoryear{Ataei et~al.}{2021}]{ataei2021theory}
\begin{barticle}
\bauthor{\bsnm{Ataei}, \binits{M.}},
\bauthor{\bsnm{Chen}, \binits{S.}},
\bauthor{\bsnm{Yang}, \binits{Z.}},
\bauthor{\bsnm{Peyghami}, \binits{M.R.}}:
\batitle{Theory and applications of financial chaos index}.
\bjtitle{Physica A: Statistical Mechanics and its Applications}
\bvolume{580},
\bfpage{126160}
(\byear{2021})
\end{barticle}
\endbibitem

\bibitem[\protect\citeauthoryear{Baker et~al.}{2016}]{baker2016measuring}
\begin{barticle}
\bauthor{\bsnm{Baker}, \binits{S.R.}},
\bauthor{\bsnm{Bloom}, \binits{N.}},
\bauthor{\bsnm{Davis}, \binits{S.J.}}:
\batitle{Measuring economic policy uncertainty.}
\bjtitle{The Quarterly Journal of Economics,}
\bvolume{131}(\bissue{4}),
\bfpage{1593}--\blpage{1636}
(\byear{2016})
\end{barticle}
\endbibitem

\bibitem[\protect\citeauthoryear{Baker et~al.}{2019}]{baker2019policy}
\begin{botherref}
\oauthor{\bsnm{Baker}, \binits{S.R.}},
\oauthor{\bsnm{Bloom}, \binits{N.}},
\oauthor{\bsnm{Davis}, \binits{S.J.}},
\oauthor{\bsnm{Kost}, \binits{K.J.}}:
Policy news and stock market volatility.
Working Paper 25720,
National Bureau of Economic Research
(2019)
\end{botherref}
\endbibitem

\bibitem[\protect\citeauthoryear{Al-Thaqeb and
  Algharabali}{2019}]{al2019economic}
\begin{barticle}
\bauthor{\bsnm{Al-Thaqeb}, \binits{S.A.}},
\bauthor{\bsnm{Algharabali}, \binits{B.G.}}:
\batitle{Economic policy uncertainty: A literature review}.
\bjtitle{The Journal of Economic Asymmetries}
\bvolume{20},
\bfpage{00133}
(\byear{2019})
\end{barticle}
\endbibitem

\bibitem[\protect\citeauthoryear{Megaritis et~al.}{2021}]{megaritis2021stock}
\begin{barticle}
\bauthor{\bsnm{Megaritis}, \binits{A.}},
\bauthor{\bsnm{Vlastakis}, \binits{N.}},
\bauthor{\bsnm{Triantafyllou}, \binits{A.}}:
\batitle{Stock market volatility and jumps in times of uncertainty}.
\bjtitle{Journal of International Money and Finance}
\bvolume{113},
\bfpage{102355}
(\byear{2021})
\end{barticle}
\endbibitem

\bibitem[\protect\citeauthoryear{Brogaard et~al.}{2022}]{brogaard2022moves}
\begin{barticle}
\bauthor{\bsnm{Brogaard}, \binits{J.}},
\bauthor{\bsnm{Nguyen}, \binits{T.H.}},
\bauthor{\bsnm{Putnins}, \binits{T.J.}},
\bauthor{\bsnm{Wu}, \binits{E.}}:
\batitle{What moves stock prices? the roles of news, noise, and information}.
\bjtitle{The Review of Financial Studies}
\bvolume{35}(\bissue{9}),
\bfpage{4341}--\blpage{4386}
(\byear{2022})
\end{barticle}
\endbibitem

\bibitem[\protect\citeauthoryear{Schiatti et~al.}{2015}]{schiatti2015extended}
\begin{barticle}
\bauthor{\bsnm{Schiatti}, \binits{L.}},
\bauthor{\bsnm{Nollo}, \binits{G.}},
\bauthor{\bsnm{Rossato}, \binits{G.}},
\bauthor{\bsnm{Faes}, \binits{L.}}:
\batitle{Extended granger causality: a new tool to identify the structure of
  physiological networks.}
\bjtitle{Physiological Measurement,}
\bvolume{36}(\bissue{4}),
\bfpage{827}
(\byear{2015})
\end{barticle}
\endbibitem

\bibitem[\protect\citeauthoryear{Faes and Sameshima}{2014}]{faes2014assessing}
\begin{bchapter}
\bauthor{\bsnm{Faes}, \binits{L.}},
\bauthor{\bsnm{Sameshima}, \binits{K.}}:
\bctitle{Assessing connectivity in the presence of instantaneous causality.}
In: \bbtitle{Methods in Brain Connectivity Inference Through Multivariate Time
  Series Analysis},
pp. \bfpage{87}--\blpage{112}.
\bpublisher{Taylor \& Francis}, \blocation{???}
(\byear{2014})
\end{bchapter}
\endbibitem

\bibitem[\protect\citeauthoryear{Guidi et~al.}{2011}]{guidi2011weak}
\begin{barticle}
\bauthor{\bsnm{Guidi}, \binits{F.}},
\bauthor{\bsnm{Gupta}, \binits{R.}},
\bauthor{\bsnm{Maheshwari}, \binits{S.}}:
\batitle{Weak-form market efficiency and calendar anomalies for eastern europe
  equity markets}.
\bjtitle{Journal of Emerging Market Finance}
\bvolume{10}(\bissue{3}),
\bfpage{337}--\blpage{389}
(\byear{2011})
\end{barticle}
\endbibitem

\bibitem[\protect\citeauthoryear{Woo et~al.}{2020}]{woo2020review}
\begin{barticle}
\bauthor{\bsnm{Woo}, \binits{K.-Y.}},
\bauthor{\bsnm{Mai}, \binits{C.}},
\bauthor{\bsnm{McAleer}, \binits{M.}},
\bauthor{\bsnm{Wong}, \binits{W.-K.}}:
\batitle{Review on efficiency and anomalies in stock markets}.
\bjtitle{Economies}
\bvolume{8}(\bissue{1}),
\bfpage{20}
(\byear{2020})
\end{barticle}
\endbibitem

\bibitem[\protect\citeauthoryear{Charfeddine and
  Khediri}{2016}]{charfeddine2016time}
\begin{barticle}
\bauthor{\bsnm{Charfeddine}, \binits{L.}},
\bauthor{\bsnm{Khediri}, \binits{K.B.}}:
\batitle{Time varying market efficiency of the gcc stock markets}.
\bjtitle{Physica A: Statistical Mechanics and its Applications}
\bvolume{444},
\bfpage{487}--\blpage{504}
(\byear{2016})
\end{barticle}
\endbibitem

\bibitem[\protect\citeauthoryear{Ito et~al.}{2016}]{ito2016evolution}
\begin{barticle}
\bauthor{\bsnm{Ito}, \binits{M.}},
\bauthor{\bsnm{Noda}, \binits{A.}},
\bauthor{\bsnm{Wada}, \binits{T.}}:
\batitle{The evolution of stock market efficiency in the us: a non-bayesian
  time-varying model approach}.
\bjtitle{Applied Economics}
\bvolume{48}(\bissue{7}),
\bfpage{621}--\blpage{635}
(\byear{2016})
\end{barticle}
\endbibitem

\bibitem[\protect\citeauthoryear{Ataei et~al.}{2020}]{ataei2020time}
\begin{barticle}
\bauthor{\bsnm{Ataei}, \binits{M.}},
\bauthor{\bsnm{Chen}, \binits{S.}},
\bauthor{\bsnm{Yang}, \binits{Z.}},
\bauthor{\bsnm{Peyghami}, \binits{M.R.}}:
\batitle{Time-homogeneous top-k ranking using tensor decompositions}.
\bjtitle{Optimization Methods and Software}
\bvolume{35}(\bissue{6}),
\bfpage{1119}--\blpage{1143}
(\byear{2020})
\end{barticle}
\endbibitem

\bibitem[\protect\citeauthoryear{Barnett and
  Bossomaier}{2012}]{barnett2012transfer}
\begin{barticle}
\bauthor{\bsnm{Barnett}, \binits{L.}},
\bauthor{\bsnm{Bossomaier}, \binits{T.}}:
\batitle{Transfer entropy as a log-likelihood ratio.}
\bjtitle{Physical Review Letters,}
\bvolume{109}(\bissue{13}),
\bfpage{138105}
(\byear{2012})
\end{barticle}
\endbibitem

\bibitem[\protect\citeauthoryear{Kirchg{\"a}ssner
  et~al.}{2012}]{kirchgassner2012introduction}
\begin{bbook}
\bauthor{\bsnm{Kirchg{\"a}ssner}, \binits{G.}},
\bauthor{\bsnm{Wolters}, \binits{J.}},
\bauthor{\bsnm{Hassler}, \binits{U.}}:
\bbtitle{Introduction to Modern Time Series Analysis.}
\bpublisher{Springer}, \blocation{???}
(\byear{2012})
\end{bbook}
\endbibitem

\bibitem[\protect\citeauthoryear{Bland and Altman}{1995}]{bland1995multiple}
\begin{barticle}
\bauthor{\bsnm{Bland}, \binits{J.M.}},
\bauthor{\bsnm{Altman}, \binits{D.G.}}:
\batitle{Multiple significance tests: the bonferroni method.}
\bjtitle{The British Medical Journal,}
\bvolume{310}(\bissue{6973}),
\bfpage{170}
(\byear{1995})
\end{barticle}
\endbibitem

\bibitem[\protect\citeauthoryear{Kleinberg}{1999}]{kleinberg1999authoritative}
\begin{barticle}
\bauthor{\bsnm{Kleinberg}, \binits{J.M.}}:
\batitle{Authoritative sources in a hyperlinked environment.}
\bjtitle{Journal of Association for Computing Machinery,}
\bvolume{46}(\bissue{5}),
\bfpage{604}--\blpage{632}
(\byear{1999})
\end{barticle}
\endbibitem

\bibitem[\protect\citeauthoryear{Fisher}{1992}]{fisher1992statistical}
\begin{bchapter}
\bauthor{\bsnm{Fisher}, \binits{R.A.}}:
\bctitle{Statistical methods for research workers.}
In: \bbtitle{Breakthroughs in Statistics},
pp. \bfpage{66}--\blpage{70}.
\bpublisher{Springer}, \blocation{???}
(\byear{1992})
\end{bchapter}
\endbibitem

\end{thebibliography}

\end{document}